\documentclass[prl,twocolumn,amsmath,superscriptaddress]{revtex4-1}
\usepackage{graphicx}
\usepackage{bm}
\usepackage[usenames,dvipsnames,svgnames,table]{xcolor}
\usepackage{soul}
\newcommand{\angstrom}{\text{\normalfont\AA}}
\newcommand{\ket}[1]{|#1\rangle}
\newcommand{\bracket}[1]{\langle #1 \rangle}

\usepackage{siunitx} 
\usepackage[usenames,dvipsnames,svgnames,table]{xcolor}
\usepackage[colorlinks=true,linkcolor=RoyalBlue,citecolor=RoyalBlue]{hyperref}

\begin{document}
\title{Thermal Hall Effect Induced by Magnon-Phonon Interactions}
\author{Xiaoou Zhang}
\affiliation{Department of Physics, Carnegie Mellon University, Pittsburgh, Pennsylvania 15213, USA}

\author{Yinhan Zhang}
\affiliation{Department of Physics, Carnegie Mellon University, Pittsburgh, Pennsylvania 15213, USA}

\author{Satoshi Okamoto}
\affiliation{Materials Science and Technology Division, Oak Ridge National Laboratory, Oak Ridge, Tennessee 37831, USA}

\author{Di Xiao}
\affiliation{Department of Physics, Carnegie Mellon University, Pittsburgh, Pennsylvania 15213, USA}

\begin{abstract}
We propose a new mechanism for the thermal Hall effect in exchange spin-wave systems, which is induced by the magnon-phonon interaction.  Using symmetry arguments, we first show that this effect is quite general, and exists whenever the mirror symmetry in the direction of the magnetization is broken.  We then demonstrate our result in a collinear ferromagnet on a square lattice, with perpendicular easy-axis anisotropy and Dzyaloshinskii-Moriya interaction from mirror symmetry breaking.  We show that the thermal Hall conductivity is controlled by the resonant contribution from the anti-crossing points between the magnon and phonon branches, and estimate its size to be comparable to that of the magnon mediated thermal Hall effect.
\end{abstract}

\maketitle

The spin-lattice interaction in solids is responsible for a wide spectrum of cross-correlated phenomena.  A well-known example is the coupling between dielectric and magnetic order in multiferroics~\cite{eerenstein2006,cheong2007,tokura2014}.  It can also manifest in the dynamics of elementary excitations such as magnons and phonons, in the form of magnon-phonon interaction.  For example, it has been demonstrated that magnons that couple to optical phonons can be launched by an electric field~\cite{pimenov2006,takahashi2012}, paving the way to the electric generation of magnon spin current~\cite{chen2015}.  On the other hand, the dynamics of phonons can be modified by the magnon-phonon interaction as well, as in the case of nonreciprocal sound propagations observed in Cu$_2$OSeO$_3$ with an applied magnetic field~\cite{nomura2018}.

Another scenario in which the magnon-phonon interaction is expected to play a significant role is the thermal Hall effect.
In a magnetic insulator, the heat current can be carried by either magnons or phonons.  Thus the thermal Hall effect can be used as an effective probe of these charge-neutral excitations.  Indeed, thermal Hall effects attributed to magnons~\cite{Onose2010,Hirschberger2015,hirschberger2015prl,gao2019} and phonons~\cite{strohm2005,Ideue2017,sugii2017,Zhang2010_2} have been reported.  Theoretical explanations have so far assumed that the low-energy excitations can be described by independent magnons or phonons~\cite{katsura2010,matsumoto2011a,Matsumoto2011,sheng2006,kagan2008,Qin2012,Mori2014,lee2015}.  However, if their interaction is strong, considering the magnon-phonon hybrid as a whole is more appropriate.  Recently, Takahashi and Nagaosa have studied the magnon-phonon interaction arising from long-range dipolar couplings~\cite{Takahashi2016}. However, the consequence of the magnon-phonon interaction from short-range couplings (such as symmetric or antisymmetric exchange) on the thermal Hall effect is yet to be explored.

\begin{figure}[t]
  \centering
  \includegraphics[width=\columnwidth]{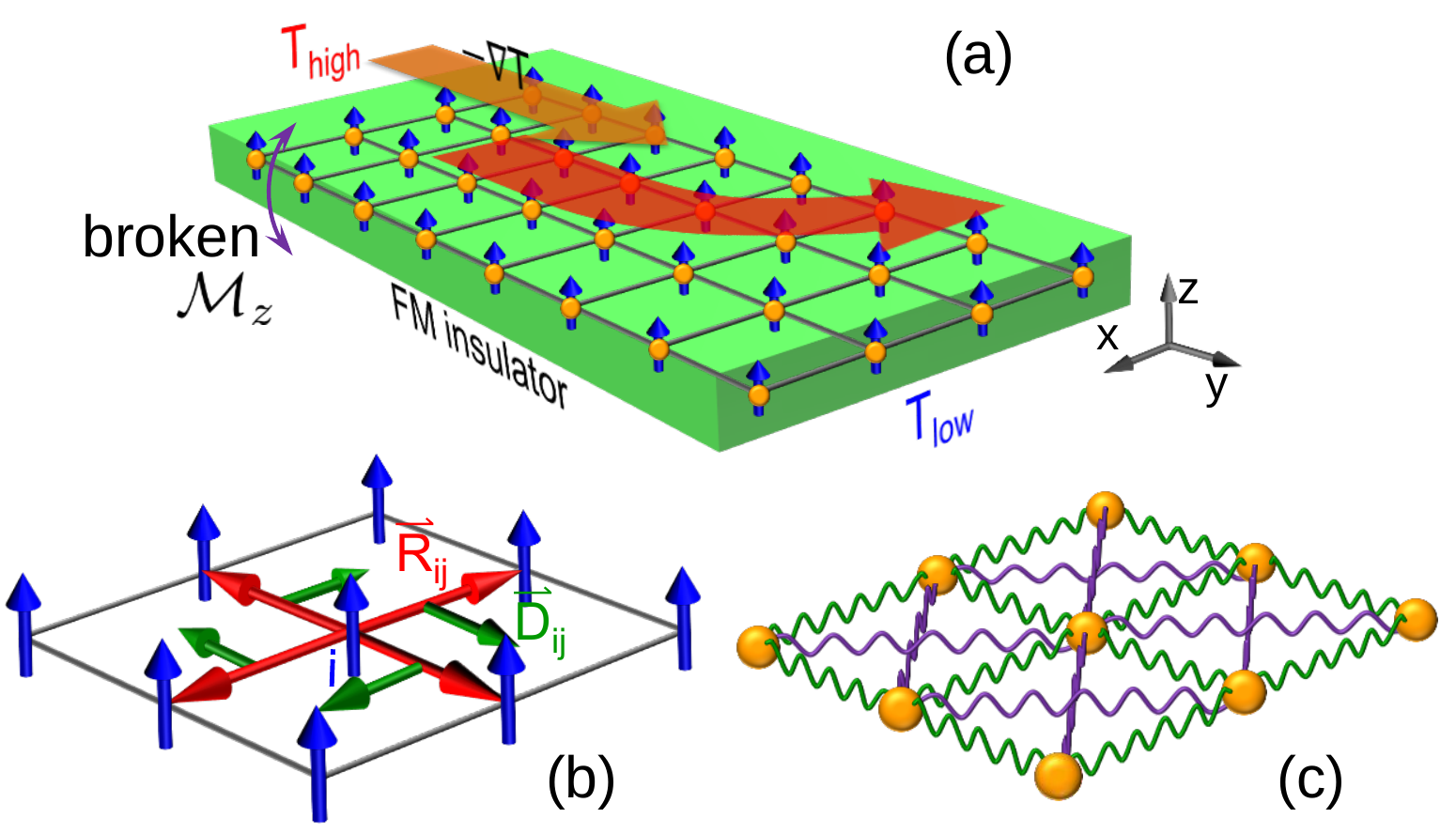}
  \caption{(a) The setup illustrates the thermal hall effect of the hybrid magnon-phonon system.  Note that the out-of-plane mirror symmetry is broken. (b) For the spin system, the ferromagnetic Heisenberg exchange interaction and the anisotropy develop a collinear ferromagnetic state with an out-of-plane magnetization (blue arrow), and the out-of-plane mirror symmetry breaking produces an in-plane DM interaction (green arrow), perpendicular to the nearest-neighbor bond direction (red arrow); (c) For the phonon system, an idealized lattice vibration model with the first (green wavy line) and second nearest neighbor interaction (purple wavy line) are considered.}
\label{device}
\end{figure}

In this Letter we investigate the effect of the magnon-phonon interaction on the thermal Hall effect.  Using symmetry arguments, we show that the magnon-phonon interaction can induce a thermal Hall effect whenever the mirror symmetry in the direction of the magnetization is broken.  In the limit of strong magnetic anisotropy, this effect can be understood as a phonon Hall effect, driven by an effective magnetic field in the phonon sector induced by the magnon-phonon interaction.  In the more general case where the magnons and phonons are close in energy, we have developed a theory to treat both excitations on an equal footing.  We demonstrate our theory in a collinear ferromagnet on a square lattice, with perpendicular easy-axis anisotropy and Dzyaloshinskii-Moriya (DM) interaction from mirror symmetry breaking (Fig.~\ref{device}).  In this model, the thermal Hall effect is entirely due to the magnon-phonon interaction.  We find that the thermal Hall conductivity is controlled by the resonant contribution from the anti-crossing points between the magnon and phonon branches, and estimate its size to be comparable to that of the magnon mediated thermal Hall effect.  Our result sheds new light on the dynamical aspect of the spin-lattice interaction, and may find applications in the emerging field of spin caloritronics~\cite{bauer2012}.

\textit{Symmetry consideration.}---We begin our discussion by analyzing the symmetry of a magnon-phonon coupled system. Consider a two-dimensional (2D) spin system described by the Hamiltonian
\begin{equation}
H_{\text{s}}=-J\sum_{\bracket{i,j}}\bm s_i\cdot\bm s_j-\frac{K}{2}\sum_i s_{iz}^2+\sum_{\bracket{i,j}}\bm D_{ij}\cdot(\bm s_i\times\bm s_j)\;,
\label{magnon}
\end{equation}
where $J>0$ represents the nearest-neighbor ferromagnetic exchange, and $K>0$ is the perpendicular easy-axis anisotropy.  The third term describes the DM interaction due to the out-of-plane mirror-symmetry breaking~\cite{fert1980,fert2013}. Here $\bm D_{ij}=D\hat{\bm R}_{ij}\times\hat{z}$ with $D$ being the strength of the DM interaction, and $\hat{\bm R}_{ij}\equiv(\bm R_{i}-\bm R_{j})/(|\bm R_{i}-\bm R_{j}|)$ is the bond direction from site $j$ to site $i$.  The direction of $\bm D_{ij}$ is in-plane and perpendicular to the bond direction, as shown in Fig.~\ref{device}(b).  We restrict our discussion to $D<\sqrt{JK}/2$ such that the ground state remains a collinear ferromagnet~\cite{Banerjee2014,supp}.

The spin-wave Hamiltonian can be obtained by expanding the spin operator in Eq.~\eqref{magnon} around its ground state expectation value, i.e., $\delta \bm s_i = \bm s_i - S\hat{z}$.  To the lowest order, the linearized spin-wave Hamiltonian reads
\begin{equation}\label{lsw}
H_{\text{sw}}=-J\sum_{\langle i,j\rangle}\delta \bm s_{i\perp}\cdot\delta \bm s_{j\perp}-(J\zeta + K) S\sum_{i}\delta s_{iz}\;,
\end{equation}
where $\zeta$ is the coordination number.  Note that the DM interaction is absent.  This can be seen by expanding the DM interaction,
\begin{equation}
\label{DMI}
H_\text{DMI}=DS\sum_{\langle i,j\rangle}\hat{\bm R}_{ij}\cdot(\delta\bm s_{i}-\delta\bm s_{j})+\mathcal{O} (\delta s^3)\;.
\end{equation}   
After summing over all lattice sites, the total DM interaction vanishes within the linear spin-wave theory.  This is a general consequence of the DM vector $\bm D_{ij}$ being perpendicular to the magnetization. If $\bm D_{ij}$ is parallel to the magnetization, then the DM interaction explicitly enters into the spin-wave Hamiltonian and, as shown in previous work, gives rise to a thermal Hall effect carried by magnons~\cite{Onose2010,hirschberger2015prl,katsura2010,matsumoto2011a,Matsumoto2011}.

Since the DM interaction is absent in the spin-wave Hamiltonian, the magnon subsystem alone does not exhibit the thermal Hall effect.  This can also be understood by the following symmetry consideration.  The thermal Hall effect is described by $\bm j^Q = \alpha_{xy} \hat{\bm z} \times \bm\nabla T$, where $\bm j^Q$ is the heat current, $\bm\nabla T$ is the temperature gradient, and $\alpha_{xy}$ is the thermal Hall conductivity.  Even though the spin-wave Hamiltonian in Eq.~\eqref{lsw} breaks the time-reversal symmetry, it remains invariant under the combined time-reversal ($\mathcal T$) and spin rotation ($\mathcal C_x$) by \ang{180} around the $x$-axis (or any in-plane axis).  Since $\bm j^Q$ is odd and $\bm\nabla T$ is even under $\mathcal{TC}_x$,  the existence of the $\mathcal{TC}_x$ symmetry forbids the thermal Hall effect.  This is reminiscent of a well-known fact about the anomalous Hall effect: it vanishes in a uniform ferromagnet in the absence of the spin-orbit interaction~\cite{gosalbez-martinez2015}.

For the phonon part, we consider a simple coupled-oscillator model described by the Hamiltonian
\begin{equation}\label{phonon}
H_{\text{ph}}=\sum_i \frac{\bm p_i^2}{2M}+\frac{1}{2}\sum_{i,j,\alpha,\beta} u^{\alpha}_i \Phi^{\alpha \beta}_{ij}u^{\beta}_j\;,
\end{equation}
where $M$ is the ion mass, $\bm u_i \equiv\bm R_i-\bm R_i^0 $ is the displacement of the $i$th ion from its equilibrium position $\bm R_i^0$, $\bm p_i = \dot{\bm u}_i$ is the canonical momentum conjugate to $\bm u_i$, and $\Phi^{\alpha \beta}_{ij}$ is the dynamical matrix describing inter-ion interactions. Obviously, due to the presence of time-reversal symmetry, the phonon subsystem alone does not exhibit the thermal Hall effect either.

The magnon-phonon interaction enters through the dependence of the exchange interaction on the ion displacement $\bm u_i$, i.e., phonons.  For the isotropic Heisenberg exchange, we find that expanding $J(R_{ij})$ in terms of $\bm u_i$ only normalizes the magnon energy, and cannot lead to the thermal Hall effect since it preserves the rotational symmetry in the spin space~\cite{Takahashi2016}. On the other hand, the in-plane DM interaction will have a nontrivial contribution to the magnon-phonon hybrid.  Expanding the DM interaction in Eq.~\eqref{DMI} to the first order in $\bm u_i$, we find
\begin{equation}
H_{\text{int}}=\sum_{\bracket{i,j}}\sum_{\alpha,\beta}(u^{\alpha}_{i}-u^{\alpha}_{j})T^{\alpha\beta}({\bm R}^{0}_{ij})(\delta s^{\beta}_i-\delta s^{\beta}_j)\;,
\label{dm_expand}
\end{equation}
where $T^{\alpha\beta}(\bm R)$ is the magnon-phonon coupling matrix,
\begin{equation} \label{TTT}
T^{\alpha\beta}(\bm R) = \frac{D}{|\bm R|}S[\delta^{\alpha\beta}-(1+\gamma)\hat{R}^{\alpha}\hat{R}^{\beta}] \;,
\end{equation}
with $\gamma = -(dD/dR)/(D/R)$.  In obtaining Eq.~\eqref{TTT}, we note that the DM interaction depends on both the bond length $R_{ij}$ and the bond direction $\hat R_{ij}$.  It is clear that at the lowest order of the expansion, only the in-plane phonon modes are involved in the magnon-phonon interaction, and we shall only consider these modes from now on.  

Since the magnon-phonon interaction in Eq.~\eqref{dm_expand} couples the spin $\delta\bm s$ to the displacement field $\bm u$, it can be regarded as an effective spin-orbit interaction for the magnon-phonon hybrid.  In particular, it breaks the $\mathcal{TC}_x$ symmetry, making the thermal Hall effect possible.  We have therefore found an interesting example in which neither the magnons nor the phonons alone exhibit the thermal Hall effect, but the magnon-phonon hybrid could via the magnon-phonon interaction.  

We can also deduce the dependence of the thermal Hall conductivity $\alpha_{xy}$ on the DM interaction $\bm D_{ij}$ and the magnetization $\bm M$.  Since $\alpha_{xy}$ is invariant under the out-of-plane mirror reflection, flipping the sign of $D$, which is determined by the mirror-symmetry breaking, does not change the sign of $\alpha_{xy}$, i.e., $\alpha_{xy}$ must be an even function of $D$.  However, if we flip the direction of the magnetization $\bm M$, the whole system turns into its time-reversal counterpart. Therefore, reversing the ground state magnetization changes the sign of $\alpha_{xy}$.

\textit{Large magnetic anisotropy limit.}---Having established the symmetry requirement for the magnon-phonon interaction induced thermal Hall effect, we now develop a quantitative theory.  Let us first consider the limit of large magnetic anisotropy, $K \gg k_BT$.  In this limit, the magnons are pushed well above phonons in energy, and the thermal transport is mainly contributed by phonons.  We can thus integrate out the magnon degree of freedom to obtain an effective Hamiltonian for phonons~\cite{liu2017}.  Leaving the details in the Supplementary Material~\cite{supp}, we find that the effective Hamiltonian for phonons is given by
\begin{equation} \label{heff}
H_{\text{ph}}^{\text{eff}}=\sum_{\bm q}\frac{(\bm p_{-\bm q}-\bm{\mathcal{A}}_{-\bm q}\bm u_{-\bm q})^T(\bm p_{\bm q}-\bm{\mathcal{A}}_{\bm q}\bm u_{\bm q})}{2M}+\frac{1}{2}{\bm u_{-\bm q}^T \bm\phi_{\bm q}\bm u_{\bm q}}\;,
\end{equation}
where ${\bm\phi}_{\bm q}^{\alpha\beta}\equiv{{\Phi}}_{\bm q}^{\alpha\beta}+\delta {{\Phi}}_{\bm q}^{\alpha\beta}$ is the renormalized dynamical matrix, and ${\mathcal{A}}_{\bm q}^{\alpha\beta}$ is the emergent gauge field experienced by phonons. Detailed calculation shows that ${\delta{\Phi}}_{\bm q}^{\alpha\beta}$ and ${\mathcal{A}}_{\bm q}^{\alpha\beta}$ are proportional to the real part and the imaginary part of the spin-spin response function of the ferromagnetic state, respectively~\cite{supp}.  Eq.~\eqref{heff} describes a phonon system in a perpendicular magnetic field~\cite{wang2009}, and can lead to the thermal Hall effect of phonons.

We note that the mechanism of this phonon Hall effect is different from that originated from the Raman type spin-lattice interaction~\cite{capellmann1991,sheng2006}. In the Raman type interaction, the phonon modes couple to the static spin ground state, while in our model, phonons couple to magnons which describe the dynamic of the spin system.

\textit{Magnon-phonon hybrid.}---If the magnon and phonon bands are close in energy, we need to treat them  on an equal footing and consider the complete Hamiltonian that includes both magnons and phonons, i.e., $H=H_{sw}+H_{\text{ph}}+H_{\text{int}}$. As a simple example, we consider a magnon-phonon interacting system on a 2D square lattice.  The linear spin wave model in Eq.~\eqref{lsw} can be solved by applying the Holstein-Primakoff transformation~\cite{holstein1940}, $\delta s_{ix}=\sqrt{S/2}(a_i+a_i^\dag)$, $\delta s_{iy}=-i\sqrt{S/2}(a_i-a_i^\dag)$, and $\delta s_{iz}=-a_i^\dag a_i$, where $a_i$ and $a_i^\dag$ are the creation and annihilation operators for magnons at the $i$-site.  This transformation gives rise to the magnon band dispersion $E_{m\bm q}=2SJ[2-\cos(q_xa)-\cos(q_ya)]+{K}(2S-1)/2$. For the phonon part, we consider the first and the second nearest neighbor interactions (see Fig.~\ref{device}(c)). The dynamic matrix in this case is given in the Supplementary Material~\cite{supp}.

The dynamics of the magnon-phonon hybrid excitation can be determined by the generalized Bogoliubov-de Gennes (BdG) equation.  To this end, we transform into the Fourier space and work in the basis of $\hat{\bm\psi}_{\bm q}=[(a_{\bm q}+a_{-\bm q}^\dag)/\sqrt{2},(a_{\bm q}-a_{-\bm q}^\dag)/(\sqrt{2}i),\tilde{u}_{\bm q}^x,\tilde{u}_{\bm q}^y,\tilde{p}_{-\bm q}^x,\tilde{p}_{-\bm q}^y]^T$, where the dimensionless operators are given by $\tilde{u}_{\bm q}^\alpha=\sqrt{M\Omega/\hbar}{u}_{\bm q}^\alpha$ and $\tilde{p}_{\bm q}^\alpha=\sqrt{1/M\Omega\hbar}{p}_{\bm q}^\alpha$, and $\Omega$ is the vibration frequency of nearest neighbor ions. From the Heisenberg equation of motion $i\hbar\partial_t\hat{\bm\psi}_{\bm q}=[\hat{\bm\psi}_{\bm q},H]$, we obtain~\cite{supp}
\begin{equation}\label{eom_hybrid}
i\hbar\mathcal{J}\partial_t\hat{\bm\psi}_{\bm q}=\mathcal{H}_{\bm q}\hat{\bm\psi}_{\bm q} \;,
\end{equation}
where the matrix $\mathcal J$ is given by
\begin{equation}
\mathcal{J}=[\hat{\bm\psi}_{\bm q},\hat{\bm\psi}_{\bm q}^\dag]=\begin{pmatrix}
-\sigma_y & 0 & 0\\
0 & 0 & iI_{2\times2}\\
0 & -iI_{2\times2} & 0
\end{pmatrix}\;.
\end{equation}
The effective Hamiltonian matrix of the hybrid system $\mathcal{H}_{\bm q}$ has the form
\begin{equation}\label{H_eff}
\mathcal{H}_{\bm q}=\begin{pmatrix}
E_{m \bm q}I_{2\times2} & \bm M_1^\dag & 0\\
\bm M_1 & \tilde{\bm\Phi}(\bm q) & 0\\
0 & 0 & \hbar\Omega I_{2\times2}\\
\end{pmatrix}\;,
\end{equation}
where 
$\bm M_1$ is a real diagonal matrix proportional to the DM strength $D$, given by
\begin{equation}
\begin{aligned}
\bm M_1=&\frac{D}{a}\sqrt{\frac{\hbar S^3}{2M\Omega}}\begin{pmatrix}
2-2\cos(q_ya) & 0\\
0 & 2-2\cos(q_xa)\\
\end{pmatrix}\\
&-\frac{\gamma D}{a}\sqrt{\frac{\hbar S^3}{2M\Omega}}\begin{pmatrix}
2-2\cos(q_xa) & 0\\
0 & 2-2\cos(q_ya)\\
\end{pmatrix}\;,
\end{aligned}
\end{equation}
and $\tilde{\bm\Phi}(\bm q)=\hbar\bm\Phi(\bm q)/(M\Omega)$.

The frequency of the magnon-phonon hybrid excitation can be derived by solving the eigenvalue problem of the generalized BdG equation $\mathcal{E}_{n\bm q}\mathcal{J}\Psi_{n\bm q}=\mathcal{H}_{\bm q}\Psi_{n\bm q}$.
Note that this system has a particle-hole symmetry, meaning that the spectrum has a positive branch and a negative branch. Since the excitation spectrum can only have positive energies, the negative branch is redundant.

We derive the thermal Hall conductance of the magnon-phonon hybrid excitation using the wave packet theory~\footnote{we have proved elsewhere that the result using the wave packet theory is the same as that from the linear response theory.}. The wave packet of the hybrid excitation is written as $\ket{W}=\int dq^2 w(\bm q, t)e^{i\bm q\cdot \bm r}\ket{\Psi_{n\bm q}}$, where $w(\bm q, t)$ is the envelop function centered around the center-of-mass momentum $\bm q_c=\int dq^2 |w(\bm q, t)|^2 \bm q$. Accordingly, the center of the wave packet in real space is given by $\bm r_c=\bracket{W|\mathcal{J}{\bm r}|W}/\bracket{W|\mathcal{J}|W}$. We derive the equation of motion for the wave packet from the Lagrangian $\mathcal{L}=\bracket{W|i\hbar\mathcal{J}d/dt-\mathcal{H}|W}/\bracket{W|\mathcal{J}|W}$, and $\bracket{W|\mathcal{H}|W}/\bracket{W|\mathcal{J}|W}=\mathcal{E}_{n\bm q_c}$. Using the Euler-Lagrangian equation, we can obtain the equation of motion for the wave packet center $\bm r_c$~\cite{ran_prl2016,Qin2012,zhang2018a,zhang2010}
\begin{equation}
\dot{\bm r_c}=\frac{\partial\mathcal{E}_{n\bm q_c}}{\hbar \partial\bm q_c}+\frac{1}{\hbar}\nabla U(\bm r_c)\times\bm \Omega_n(\bm q_c) \;,
\end{equation}
where  $U(\bm r)$ is the potential felt by the wave packet, and the Berry curvature is defined by $\Omega^{z}_n=\partial_{q_x} A_{ny}-\partial_{q_y} A_{nx}$, with $\bm A_n=i\bracket{\Psi_{n\bm q}|\mathcal{J}\partial_{\bm q}|\Psi_{n\bm q}}/\bracket{\Psi_{n\bm q}|\mathcal{J}|\Psi_{n\bm q}}$ being the Berry connection.

\begin{figure}[t]
  \centering
  \includegraphics[width=1.0\columnwidth]{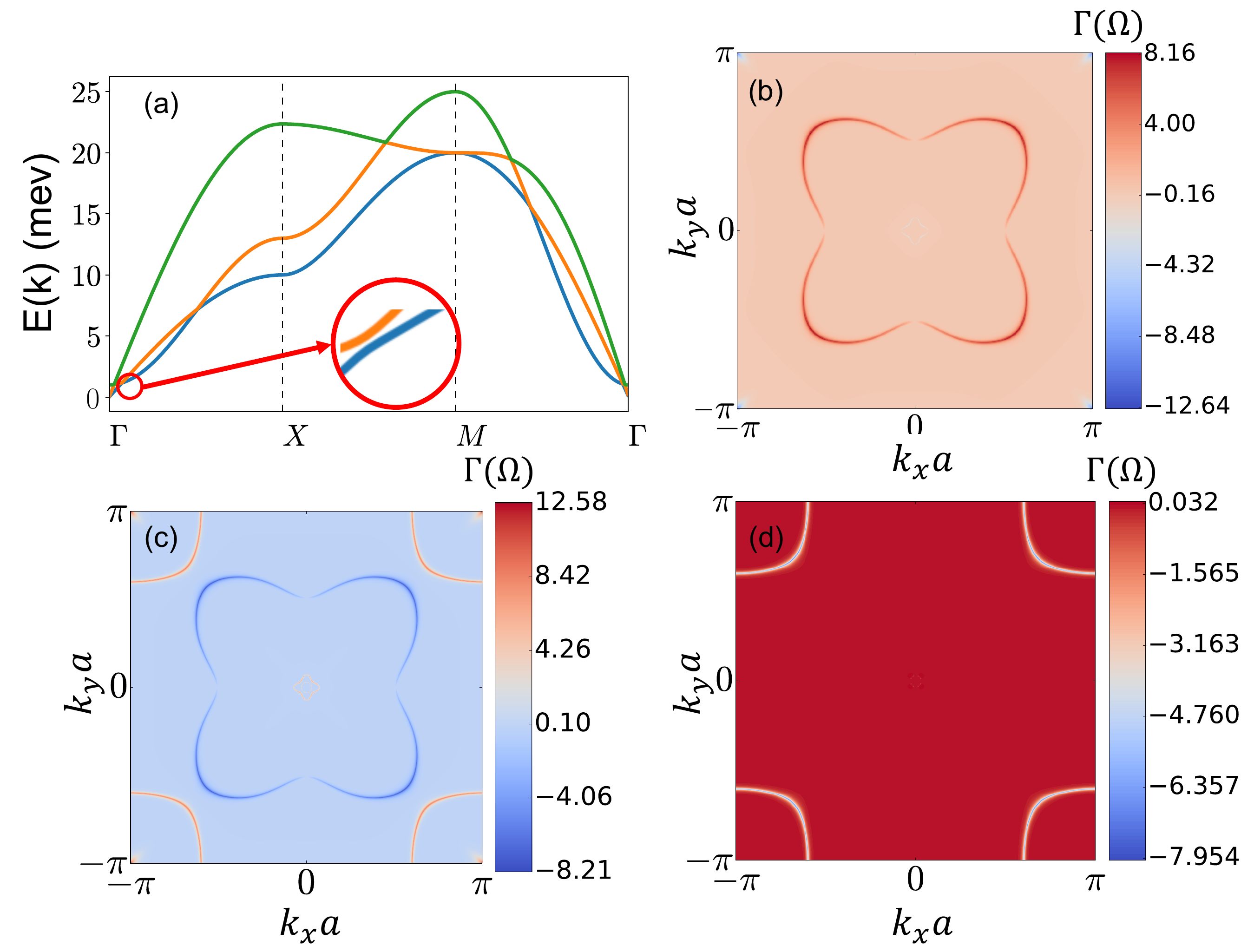} 
  \caption{The band structure and Berry curvature using the parameters in the main text with $D = 0.4$ meV. (a) The band structure of the magnon-phonon hybrid system along the high symmetry line $\Gamma-X-M-\Gamma$. The degeneracies in the bands are lifted by the magnon-phonon interaction, as shown in the inset; (b,c,d)~The distribution of Berry curvatures in log-scale $\Gamma(\Omega^z) \equiv\text{sign}(\Omega^z)\ln(1+|\Omega^z|)$ for (b) the lowest band, (c) the middle band, and (d) the highest band.}
\label{band_crossing}
\end{figure} 

Using the equation of motion of the wave packet, the thermal Hall current for Bosonic excitations is given by~\cite{sundaram1991,di_prl2006,Matsumoto2011,matsumoto2011a}:
\begin{equation} \label{alpha}
\begin{aligned}
\bm j&=\frac{k_B^2T}{\hbar}\hat{\bm z}\times \nabla T\sum_n\int \frac{d^2q}{(2\pi)^2}\Omega^{z}_n(\bm q)\\
&\Big[(1+\rho_{n\bm q})\ln^2\Big(\frac{1+\rho_{n\bm q}}{\rho_{n\bm q}}\Big)-\ln^2 \rho_{n\bm q}-2\text{Li}_2(-\rho_{n\bm q})\Big]\;.
\end{aligned}
\end{equation}
Here $\rho_{n\bm q}=1/(e^{E_n(\bm q)/k_BT}-1)$ is the Bose-Einstein distribution function with a zero chemical potential, and the index $n$ in the Berry curvature $\Omega^{z}_n$ is summed over all positive bands. 

\textit{Berry-curvature hotspots}.---A generic feature of the magnon-phonon hybrid bands is the existence of anti-crossing points due to the magnon-phonon interaction.  These anti-crossing points give rise to the Berry-curvature hotspots that contribute resonantly to the thermal Hall conductivity and lead to a sizable effect.  To demonstrate this, below we carry out numerical estimation of the thermal Hall conductivity.

\begin{figure}
  \includegraphics[width=0.75\columnwidth]{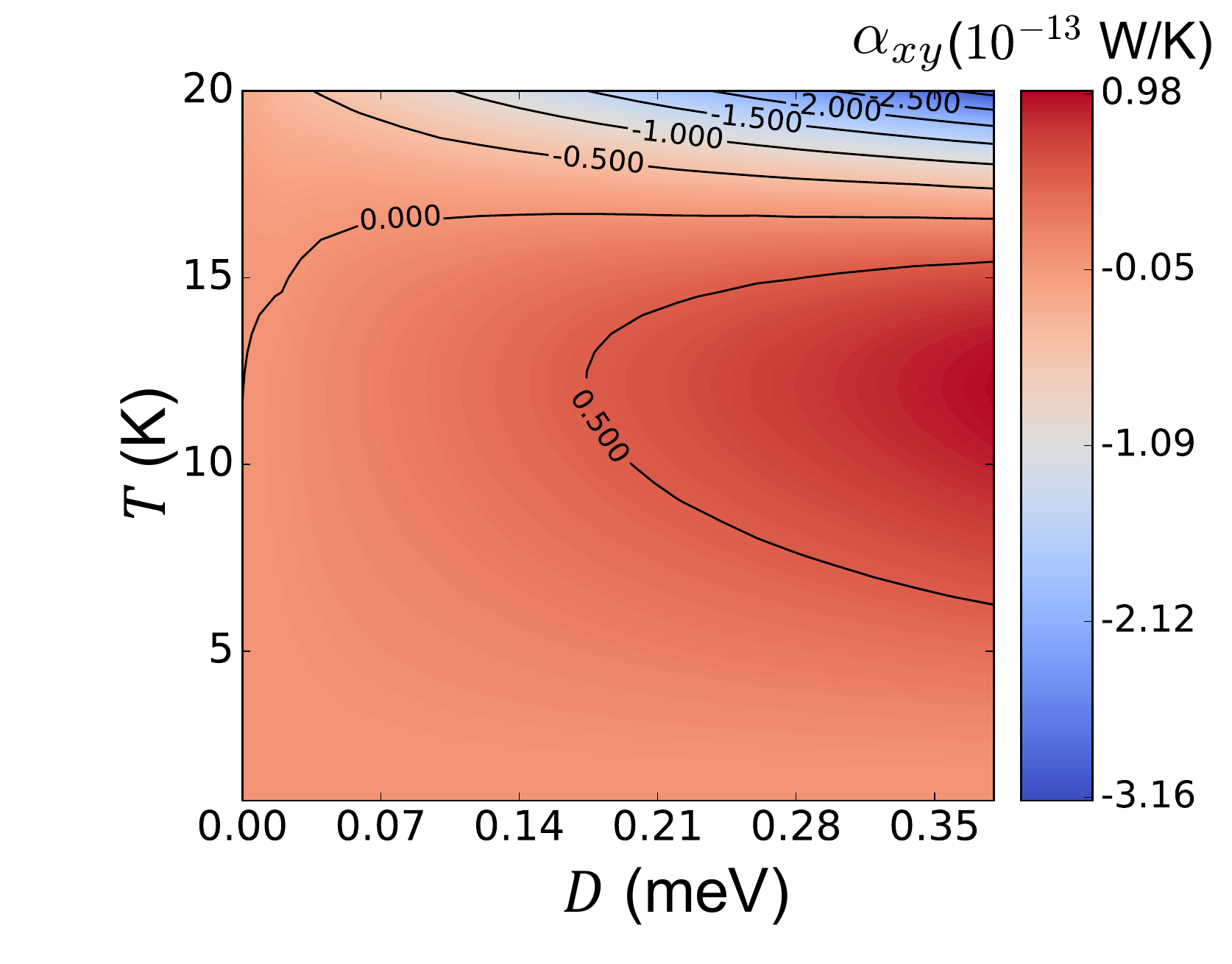}
  \caption{The thermal Hall conductivity $\alpha_{xy}$ as a function of temperature and the strength of DM interaction $D$. Other parameters are defined in the main text.}
\label{cond_contour}
\end{figure}

Suppose the magnetic ions are one of the 3$d$ transition metal atoms.  For an estimation we set the total spin $S = 3/2$ and atomic mass $M = 50$ proton mass.  The typical values of the Heisenberg exchange is on the order of meV, and we have chosen $J = 2$ meV.  For the perpendicular magnetic anisotropy, we set $K = 1$ meV, which is attainable in low-dimensional systems~\cite{huang2017}.  The most important parameter is the DM interaction due to mirror symmetry breaking.  It has been shown that DM interaction of this type can be as large as 20\% of the Heisenberg exchange $J$ in heterostructures~\cite{fert2013}.  For the phonon part, we will set the lattice vibration frequency for the nearest neighbor interactions at $10$ meV, and for the second nearest neighbor interactions at $5$ meV. For simplicity, we have set $\gamma = 0$ in Eq.~\eqref{TTT}~\footnote{In the superexchange model, the DM strength $D\sim t_0t'/U$, where $t_0$ and $t'$ are the spin-independent and spin-dependent hopping integrals and $U$ is the onsite Coulomb interaction~\cite{Moriya1960}.  In general, the hopping integrals have a power law dependence on the interatomic distance $R$~\cite{Harrison}, and the dependence of $U$ on $R$ can be neglected.  If $D \sim R^{-n}$ then $dD/dR = -nD/R$.  Therefore $\gamma = -(dD/dR)/(D/R) = n$ should have the same order as unity, and our numerical calculation based on $\gamma = 0$ should give an order of magnitude estimation of the thermal Hall conductivity.}.

Figure~\ref{band_crossing}(a) shows the band structure of the magnon-phonon hybrid.  The bands have several anti-crossing points due to the magnon-phonon interaction.  Those gaps are too small to be seen, but the Berry-curvature hotspots shown in Fig.~2(b)-(d) are their fingerprints---we have verified that those hotspots are precisely where the anti-crossing points are located.  The Berry-curvature hotspots dominate the contribution to the thermal Hall conductivity, and can lead to a large effect.  The dependence of the thermal Hall conductivity $\alpha_{xy}$ on both temperature and the strength of the DM interaction is shown in Fig~\ref{cond_contour}. 
For $T=20$ K, $D=0.2$ meV, $\alpha_{xy}\sim 1.5\times10^{-13}$ W/K. In Ref.~\cite{Onose2010}, the magnon thermal Hall conductance of the bulk sample is around $10^{-3}$ WK$^{-1}$m$^{-1}$. If we assume that the thickness of a monolayer sample is 5 $\angstrom$, then the thermal Hall conductance for one monolayer is about $5\times10^{-13}$ W/K.  Therefore, the thermal Hall conductance of our model is at the same order as that of the magnon Hall effect.  We have also verified that the order of magnitude estimation is robust against changes of the material parameters.

In summary, we have proposed a new mechanism for the thermal Hall effect in an exchange spin-wave system by magnon-phonon interactions.  The key ingredient is an out-of-plane magnetization and an in-plane DM vector due to mirror symmetry breaking.  Even though our discussion is focused on a 2D spin layer, the mirror symmetry breaking can be realized in bulk crystals consisting of stacked 2D layers with broken mirror symmetry, or in superlattices of magnetic multilayers where the mirror symmetry is broken by the heterointerface.  We note that the magnon-phonon interaction arising from long-range dipolar couplings could in principle also contribute to the thermal Hall effect~\cite{Takahashi2016}.  However, our symmetry-based mechanism can also be active in antiferromagnets where the dipolar coupling is absent.  Our result revealed the crucial role of the magnon-phonon interaction in the thermal Hall effect, and may find applications in the emerging field of spin caloritronics~\cite{bauer2012}.

We acknowledge useful discussions with Ran Cheng, Matthew W.\ Daniels, Tao Qin and Junren Shi.  Work at CMU is supported by the U.S.\ Department of Energy, Basic Energy Science, Pro-QM EFRC DE-SC0019443 (X.Z.) and DE-SC0012509 (Y.Z.\ and D.X.).  S.O.\ acknowledges support by the U.S.\ Department of Energy, Office of Science, Basic Energy Sciences, Materials Sciences, and Engineering Division.  D.X.\ also acknowledges support from a Research Corporation for Science Advancement Cottrell Scholar Award.

X.Z.\ and Y.Z.\ contributed equally to this work.

\textit{Note added} ---Upon the completion of this work, we have become aware of a recent paper~\cite{park2019} in which the thermal Hall effect from magnon-phonon interactions in noncollinear antiferromagnets is considered.

\section*{Supplementary}
\subsection{Spin Ground State}

In this section, we show that the classical ground state of the spin Hamiltonian [Eq.~(1) in the main text] remains a collinear ferromagnet for sufficiently large easy-axis anisotropy.

Let us consider a pair of nearest neighbor spins, and suppose that the angle between these two spins is given by $\Delta\theta$.  If $\Delta\theta$ is small, the exchange energy between theses two spins is given by $-{JS^2}\cos{\theta}\approx -{JS^2}(1-\frac{\theta^2}{2}) $, and the DM energy is $-{DS^2}\sin\theta\approx -{DS^2}\theta$. In the absence of anisotropy, the angle $\Delta\theta$ that minimizes the sum of the exchange and DM energy is $\Delta\theta=D/J$, which leads to a spiral phase with the period of $2\pi/\Delta\theta=2\pi J/D$. Now we consider the anisotropy energy. For a period of spiral, the energy of the spiral phase is given by
\begin{equation}
\begin{aligned}
E_{\text{spiral}}&=\frac{2\pi J}{D}\Bigl\{-JS^2+\frac{JS^2}{2}\Delta\theta^2-{DS^2}\Delta\theta \\
-& \frac{KS^2}{2}\int_0^{2\pi}\cos^2\theta \frac{d\theta}{2\pi} \Bigr\}\\
&=-\frac{2\pi J^2S^2}{D}-{\pi DS^2}-\frac{KS^2\pi J}{2D}\;.
\end{aligned}
\end{equation}
On the other hand, the energy for the collinear ferromagnetic state is given by
\begin{equation}
E_{\text{FM}}=-\frac{2\pi J^2S^2}{D}-\frac{KS^2\pi J}{D}\;.
\end{equation}
The critical value of $D$ can be determined by $E_{\text{spiral}}=E_{\text{FM}}$, which gives $D_c=\sqrt{JK/2}$.

\subsection{The Dynamic Matrix for Phonons}
Phonon modes originate from classical normal modes of vibrating ions. The Hamiltonian for ions is
\begin{equation}\label{l_phonon}
  H_{\text{ph}}=\sum_{i}\frac{{\bm p}_{i}^2}{2M}+V(\{\bm R_{i}\})\;,
\end{equation}
where $M$ is the ion's mass, and $V(\{\bm R_{i}\})$ is the potential energy between ions under some ion configuration $\{\bm{R}_i\}$. We expand the potential around the equilibrium configuration $\{\bm R^{0}_i\}$ under the harmonic approximation
\begin{equation}
\begin{aligned}
  V(\{\bm R_{i}\})\approx V(\{\bm R^{0}_{i}\})+\frac{1}{2}\sum_{i,j}\bm u_{i}\cdot\frac{\partial^2 V}{\partial \bm u_{i}\partial \bm u_{j}}\Big|_{\{\bm R_{i}=\bm R_{i}^0\}}\cdot\bm u_{j}.
\end{aligned}
\end{equation}
where the deviation from its equilibrium position of the $i$-th ion $\bm{u}_{i}\equiv \bm R_{i}-\bm R^{0}_{i}$. Here, we assume that the potential $V(\{\bm R_{i}\})$ contains only the first and second nearest neighbor interactions
\begin{equation}
\begin{aligned}
&\frac{1}{2}\sum_{i,j}\bm u_{i}\cdot\frac{\partial^2 V}{\partial \bm u_{i}\partial \bm u_{j}}\Big|_{\{\bm R_{i}=\bm R_{i}^0\}}\cdot\bm u_{j}\\
&\approx\sum_{\bracket{i,j}}\frac{M\Omega^2}{2}[(\bm u_{i}-\bm u_{j})\cdot\hat{\bm R}_{ij}^0]^2+\sum_{\bracket{\bracket{i,j}}}\frac{M\Omega'^2}{2}[(\bm u_{i}-\bm u_{j})\cdot\hat{\bm R}_{ij}^0]^2\\
&\equiv\frac{1}{2}\sum_{\bracket{i,j}}\bm{u}^{T}_i\bm\Phi_{ij}\bm{u}_j,
\end{aligned}
\end{equation} 
with two vibrations frequencies $\Omega$ and $\Omega'$ corresponding to two nearest neighbor atoms and two second nearest neighbor atoms, and $\bm{\Phi}_{ij}$ is the dynamic matrix in the real space. Here we only consider the vibrations along the bonds, since the vibrations perpendicular to the bonds are higher order effects. Accordingly, the phonon Hamiltonian can be written in the momentum $\bm q$ space
\begin{equation}
  H_{\text{ph}}=\sum_{\bm q}\frac{{\bm p}_{-\bm q}{\bm p}_{\bm q}}{2M}+\frac{1}{2}\bm u_{-\bm q}^T\bm\Phi(\bm q)\bm u_{\bm q} \;,
\end{equation}
where the dynamic matrix in the momentum space is given by
\begin{equation}
\begin{aligned}\label{dynamic_mat}
&\frac{\bm \Phi(\bm q)}{M}=2\Omega^{'2}(1-\cos{q_ya}\cos{q_xa}+\sigma_x\sin{q_xa}\sin{q_ya})\\
&+2\Omega^2\big[\sigma_z\big(\sin^2{\frac{q_xa}{2}}-\sin^2{\frac{q_ya}{2}}\big)+\sin^2{\frac{q_xa}{2}}+\sin^2{\frac{q_ya}{2}}\big]\;,
\end{aligned}
\end{equation}
where the $2\times 2$ matrices $\sigma_{x,y,z}$ are the Pauli matrices.

\subsection{Effective Phonon Model in the Square Lattice}
We build an effective theory just for phonons by integrating out the magnon degree of freedom. Our starting point is the equation of motion for ions~\cite{liu2017},
\begin{equation} \label{EoMPh}
M \ddot{ u}_i^\alpha 
= -\sum_{j,\beta} \Phi_{ij}^{\alpha\beta}u_j^\beta
+ \bracket{F_i^\alpha} \;,
\end{equation}
where $F_i^\alpha \equiv -\partial H_{\text{int}}/\partial u^\alpha_i$ is the effective force operator acting on phonons from the magnon-phonon interaction.  The expectation value $\bracket{F_i^\alpha}$ should be evaluated in the magnon subsystem subjected to a time-dependent perturbation from the lattice vibration ${\bm u_{i}}$. The bracket $\bracket{...}$ denotes the statistical quantum average of the spin states. Following Eq.~(5) in the main text, and transforming to the momentum space, we can write the effective force operator as 
\begin{equation}
\label{force}
F_{\bm q}^\alpha(t) = -\sum_{\beta}T^{\alpha\beta}_{\bm q}\delta s^{\beta}_{\bm q}(t),
\end{equation} where the coupling matrix $T^{\alpha\beta}_{\bm q}=\sum_{\bm\delta_j}(1-e^{-i\bm\delta_j\cdot\bm q})T^{\alpha\beta}(\bm \delta_j)$ with $\bm\delta_j$ the nearest neighbor vector is a symmetric matrix.  Standard linear response theory in the frequency representation can be explicitly written as~\cite{liu2017}
\begin{equation} \label{force}
\bracket{\bm{F}_{\bm q}(\omega)}=\bm{T}(\bm q)\bm{\chi}(\bm q;\omega)\bm{T}(-\bm q)\bm{u}_{\bm q}(\omega)\;,
\end{equation}
where $\chi^{\alpha\beta}(\bm q;\omega)=-\frac{i}{\hbar}\int dt e^{i\omega t}\Theta(t)\bracket{[\delta s_{\bm q}^\alpha(t),\delta s_{-\bm q}^\beta(0)]}$ is the spin-spin response function of the ferromagnetic state. Here we use the convention that a bold form such as $\bm{\chi}$, $\bm{T}$, $\bm F$ and $\bm u$ denotes a tensor, and the plain form such as $\chi^{\alpha\beta}$ denotes a tensor component.

To proceed further, let us consider the low temperature regime where only the modes with low frequencies $\omega$ are important.  Therefore, we can expand $\bm{\chi}(\bm q;\omega)$ to the first order of $\omega$, $\bm{\chi}(\bm q;\omega)\approx \bm{\chi}_0(\bm q)+i\omega \bm{\chi}_1(\bm q)$.  Inserting Eq.~\eqref{force} back to Eq.~\eqref{EoMPh}, we obtain
\begin{equation}\label{eff_ph_eom}
\Big[\Big(i\omega\bm I +\frac{\bm g({\bm q})}{2{M}}\Big)^2+\frac{\bm{\mathcal K}({\bm q})}{M}\Big]\bm u_{\bm q}=0\;, 
\end{equation}
where $\bm{\mathcal{K}}({\bm q})=\bm \Phi({\bm q})+\delta\bm{\Phi}({\bm q})-\bm{\mathcal{A}}^2({\bm q})/M$, with 
\begin{equation}\label{eff_screen}
\delta\bm{\Phi}({\bm q})=\bm{T}(\bm q)\bm{\chi_0}(\bm q)\bm{T}(-\bm q)
\end{equation}
and
\begin{equation}\label{eff_mag}
\bm{\mathcal{A}}({\bm q})=2\bm{T}(\bm q)\bm{\chi}_1(\bm q)\bm{T}(-\bm q)\;.
\end{equation}
Note that these two corrections $\delta\bm{\Phi}$ and $\bm{\mathcal{A}}$ are proportional to the real and imaginary part of the spin-spin response function, respectively. It is straightforward to show that Eq.~\eqref{eff_ph_eom} can be simply derived from the effective Hamiltonian (Eq.~(10)) in the main text. Therefore, Re${\bm{\chi}(\bm q;\omega)}$ provides a screening effect to the inter-atomic interaction, and Im${\bm{\chi}(\bm q;\omega)}$ provides an effective magnetic field for phonons.

To be specific, we calculate the $\delta\bm{\Phi}({\bm q})$ and $\bm{\mathcal{A}}({\bm q})$ terms in the square lattice. We first calculate $\bm{\chi}(\bm q;\omega)$ using the analytical continuation of its corresponding Matsubara Green's function in the frequency representation. We get 

\begin{equation}
\bm{\chi}(\bm q;\omega)=\frac{2}{\hbar(\omega^2-\omega_{m\bm q}^2)}\begin{pmatrix}
-{\omega_{m\bm q}} & i{\omega}\\
-i{\omega} & -{\omega_{m\bm q}}
\end{pmatrix}\;,
\end{equation}
where $\omega_{m\bm q}=E_m(\bm q)/\hbar$ is the magnon frequency. In the square lattice, we have $T^{xx(yy)}(\bm q)=-DS/a\{2-2\cos(q_{y(x)}a)-\gamma[2-2\cos(q_{x(y)}a)]\}$ and $T^{xy(yx)}(\bm q)=0$. 

Substituting $\bm T(\bm q)$ and $\bm{\chi}(\bm q;\omega)$ into  Eq.~\eqref{eff_mag} and \eqref{eff_screen} and gives the effective magnetic field
\begin{equation}
{\mathcal{A}}^{\alpha\beta}(\bm q)=-4\frac{c_{0}(\bm q)}{\hbar\omega_{m\bm q}^2}\epsilon^{\alpha\beta}\;,
\end{equation}
where $\epsilon$ is the antisymmetric unit tensor and $c_{0}(\bm q)=2S^3D^2[1-\cos(q_x a)-\gamma(1-\cos(q_y a))][1-\cos(q_y a)-\gamma(1-\cos(q_x a))]/a^{2}$, and the screening term
\begin{equation}
\delta\Phi^{\alpha\beta}(\bm q)=-2\delta^{\alpha\beta} \frac{c_{\alpha}(\bm q)}{\omega_{m\bm q}}\;,
\end{equation}
where $c_{x(y)}(\bm q)=2S^3D^2[1-\cos(q_{y(x)} a)-\gamma(1-\cos(q_{x(y)} a))]^2/a^{2}$.
%
Accordingly, the leading order of the screening effect $\delta\bm{\Phi}$ and the effective magnetic field $\bm{\mathcal{A}}$ from the magnon-phonon coupling is independent of the temperature. All temperature dependent terms come from higher order expansions.

\subsection{Magnon-Phonon Hybrid Hamiltonian}
After Fourier transformation $X_{i}=\sum_{\bm q}e^{i\bm q\cdot\bm R_i}X_{\bm q}/\sqrt{N}$ with $X_{i}$ denoting the magnon operator $a_{i}$, the displacement $\bm{u}_{i}$ and the momentum $\bm{p}_{i}$ for ions, the complete Hamiltonian $H=H_{sw}+H_{\text{ph}}+H_{\text{int}}$ (mentioned in the main text), can be written in the momentum space as
\begin{equation}
\begin{aligned}
H=&\sum_{\bm q}E_{m\bm q}a_{\bm q}^\dag a_{\bm q}+\frac{{\bm p}_{-\bm q}{\bm p}_{\bm q}}{2M}+\frac{1}{2}\bm u_{-\bm q}^T\bm\Phi(\bm q)\bm u_{\bm q}\\
+&\sqrt{\frac{S}{2}}\sum_{\bm{\delta},\alpha}u_{-\bm q}^\alpha(1-e^{i\bm q\cdot\bm \delta})[T^{\alpha x}({\bm\delta}) (a_{\bm q}+a_{-\bm q}^\dag)\\
&-iT^{\alpha y}({\bm\delta}) (a_{\bm q}-a_{-\bm q}^\dag)]\;,
\end{aligned}
\end{equation}
where $\bm\delta$ is the nearest neighbor vector, and $E_{m\bm q}=SJ(\zeta-\sum_{\bm{\delta}}e^{i\bm q\cdot\bm{\delta}})+K(2S-1)/2$ is magnon dispersion. Since the magnon operator $a^{\dagger}_{\bm q}$, $a_{\bm q}$, the displacement $\bm u_{\bm q}$ and the momentum $\bm p_{\bm q}$ for ions are considered on the same footing, the quadratic Hamiltonian can always be written as
\begin{equation}
H=\frac{1}{2}\sum_{\bm q}\hat{\bm\psi}_{\bm q}^\dag \mathcal{H}_{\bm q}\hat{\bm\psi}_{\bm q}\;,
\end{equation}
based on a column vector $\hat{\bm\psi}_{\bm q}=[(a_{\bm q}+a_{-\bm q}^\dag)/\sqrt{2},(a_{\bm q}-a_{-\bm q}^\dag)/(\sqrt{2}i),u_{\bm q}^x,u_{\bm q}^y,p_{-\bm q}^x,p_{-\bm q}^y]^T$, and $\mathcal{H}_{\bm q}$ is given in Eq.~(10) in the main text. 

Now we solve the Heisenberg equation of motion $i\hbar\partial_t\hat{\bm\psi}_{\bm q}=[\hat{\bm\psi}_{\bm q},H]$. Note that $\hat{\bm\psi}_{\bm q}^\dag=\hat{\bm\psi}_{-\bm q}$ and $\mathcal{H}_{-\bm q}^T=\mathcal{H}_{\bm q}$. Therefore, we have $i\hbar\partial_t\hat{\bm\psi}_{\bm q}=\mathcal{J}\mathcal{H}_{\bm q}\hat{\bm\psi}_{\bm q}$, where $\mathcal{J}=[\hat{\bm\psi}_{\bm q},\hat{\bm\psi}_{\bm q}^\dag]$ is given in the main text. Since $\mathcal{J}^2=I_{6\times 6}$ with $I_{6\times 6}$ being an $6\times6$ identity square matrix, we get $i\hbar\mathcal{J}\partial_t\hat{\bm\psi}_{\bm q}=\mathcal{H}_{\bm q}\hat{\bm\psi}_{\bm q}$.

\end{document}